\documentstyle[preprint,aps,prl]{revtex}
\begin{document}
\draft
\title{Electronic transport in a series of multiple\\
       arbitrary tunnel junctions}
\author{U.~E.~Volmar\footnote{Corresponding author. E-mail: u.volmar@rz.uni-sb.de}, U.~Weber, R.~Houbertz and U.~Hartmann}
\address{Institute of Experimental Physics, University of Saarbr\"ucken,\\
	 P.O. Box 151150, D--66041 Saarbr\"ucken, Germany}
\date{\today}
\maketitle
\begin{abstract}
Monte Carlo simulations and an analytical approach within
the framework of a semiclassical model are presented which permit
the determination of 
Coulomb blockade and single electron charging effects for
multiple tunnel junctions coupled in series.
The Coulomb gap in the I(V) curves can be expressed as a 
simple function of the capacitances in the series. 
Furthermore, the magnitude of the differential conductivity 
at current onset
is calculated in terms
of the model.
The results are discussed with respect to the number of 
junctions.
\end{abstract}
\vspace{1cm}
%
%
\narrowtext
Since the development of quantum mechanics, electron tunneling 
has been widely investigated experimentally 
\cite{giaever,wiesend} and extensively 
discussed in theory.\cite{simmons,bardeen}
The first experiments were performed on metal-insulator-metal (MIM) 
sandwiches,\cite{giaever} which permitted the study of tunneling 
phenomena and the direct correlation of such effects with the thickness 
of the potential barrier. 
There are two main theoretical approaches commonly used 
to describe tunneling phenomena. 
The first entails solving the time-independent Schr\"odinger equation 
and matching the wave functions.\cite{simmons} 
However, an exact matching of the wave functions can only be 
achieved in a one-dimensional approximation. 
The second method is known as the transfer-Hamiltonian approach,
\cite{bardeen} which generally involves a time-dependent 
Schr\"odinger equation in the second quantization formalism. 
In this approach, tunneling of the electrons between the two electrodes 
is described using a transfer Hamiltonian in addition to the 
Hamiltonians of the unperturbed electrodes.
 
For small-scaled systems a particular tunneling phenomenon 
referred to as single electron tunneling (SET)\cite{averin} can occur.
SET may be observed, for example, in a system of two electrodes 
separated by a thin insulating layer, in which a 
third electrode (e.g., a small metal particle) is embedded. 
The current through the system is controlled by a single electron
if  $k_B T\ll e^2/\left[2(C_1+C_2)\right]$, with 
$C_1$ and $C_2$ being the capacitances of the two junctions. 
A tunneling event therefore changes the total charge of the center
electrode by $\pm e$, depending on the direction of tunneling, 
and thus the electrostatic energy of the system. 
At zero temperature, tunneling is completely prohibited for  
$\vert V\vert <e/\left[2(C_1 + C_2)\right] = 
             \Delta V_{\scriptsize\mbox{gap}}/2$ 
(Coulomb blockade), 
and the transport process is dominated by charging effects. 
This means, in particular, that an increase in voltage applied to 
the double junction system leads to incremental charging, 
which might manifest itself as steps in the $I(V)$ characteristic
(Coulomb staircase)\nolinebreak.\cite{averin} 
Such $I(V)$ characteristics are usually measured at temperatures 
below 4 \nolinebreak K \nolinebreak \cite{vankampen} for certain
relations of $R_i C_i$, such as $R_1 C_1\ll R_2 C_2$
($R_i$ and $C_i$ are the tunnel resistance and capacitance
of the $i$-th junction). The theory developed so 
far only considers this low-temperature limit.

One important point in understanding SET  
is the extension of the theory to multiple tunnel junctions as well 
as to room temperature. 
Increasing the temperature to room temperature demands capacitances as 
small as $C\simeq 0.1$~aF, 
since only then does 
the electrostatic energy of charging by a single electron, 
$e^2/2C$,
exceed the thermal energy $k_B T$. 
Nowadays, such capacitances can be 
realized using small particles \cite{vankampen} or clusters 
\cite{clusters} with dimensions in the 1 to 10 nm range.

Here, a theoretical study of  $I(V)$ characteristics of 
one-dimensional chains of small capacitors coupled in series is
presented. Such
arrangements can be realized, for example, in granular films of small
metal islands or with clusters. 
An analytical approach for zero temperature in the framework of a 
semiclassical model is used to explain several features
arising in Monte Carlo simulations at finite temperatures.
It is shown that the width of the Coulomb gap is directly related to 
the number of capacitors in the series.
An expression is given for calculating the voltages 
at which steps in the $I(V)$ characteristic occur.
Additionally, the differential conductivity
at the current onset just beyond the Coulomb gap is calculated.

Figure \ref{schalt3} shows 
a system of $N$ tunnel junctions coupled in series.
In the semiclassical model, the state of each tunnel 
junction is characterized by the voltage drop across  
[Fig. \nolinebreak\ref{schalt3}~(a)].
The individual junction
 voltages can be calculated using Kirchhoff's law together 
with Gauss's law.
Due to tunneling there might be additional electrons on the center 
electrodes.
The voltage $V_k$ across the $k$-th junction with $n_j$ extra
electrons on the $j$-th electrode and an externally applied voltage 
$V$  
can be written as
\smallskip
\begin{eqnarray}
V_k(\ldots n_j\ldots,V) =
\frac{1}{C_k} 
\left(
\frac{C_1 V - C_1 e 
      \sum\limits_{m=2}^N \protect\normalsize
      \left(1/C_m \sum\limits_{j=1}^{m-1} n_j\right)}
     {1 + C_1 \sum\limits_{m=2}^N 1/C_m} 
 +  e \sum\limits_{j=1}^{k-1} n_j 
\right),
\end{eqnarray}
\smallskip
 where $C_m$ denotes the capacitance of the $m$-th tunnel junction. 
$V_k$  causes a mutual shift in the Fermi energies of the electrodes.

The tunneling rates across the junctions, which can be calculated using
Fermi's golden rule, are given by the following expression
(for tunneling through the $k$-th junction from right to left):
\begin{eqnarray}
r_k =
\int_{-\infty}^{+\infty} dE \
\frac{2 \pi}{\hbar} |T(E)|^2 
D_{k-1}(E - E_{k-1}) f(E - E_{k-1})
D_k(E-E_k) \left[ 1 - f(E-E_k) \right],
\end{eqnarray}
where $f(E)$ is the Fermi-Dirac distribution.
If we consider, as usual, the density of states $D$ near the Fermi 
level and the tunnel matrix element $T$ to be energy-independent,
that is $D_k(E-E_k)=D_k^0$ and $|T(E)|^2=|T_0|^2$, 
the tunneling rates from right to left 
($r_k$) and reverse ($l_k)$ [cf.\ Fig.\nolinebreak\ref{schalt3}(b)] are
\smallskip
\begin{equation}
r_k(\ldots n_j\ldots,V) = 
\frac{1}{e^2 R_k} ~
\frac{\Delta E_k^{\leftarrow}}{1 - \exp ({-\Delta E_k^{\leftarrow}/k_B T})}
\end{equation}
and
\begin{equation}
l_k(\ldots n_j\ldots,V) = 
\frac{1}{e^2 R_k} ~
\frac{\Delta E_k^{\rightarrow}}{1 - \exp ({-\Delta E_k^{\rightarrow}/k_B T})},
\end{equation}
\smallskip
where 
the tunneling resistances $R_k$ are given by 
$1/R_k = \left(2 \pi e^2/\hbar\right) \ D_{k-1}^0 D_k^0 |T_0|^2$.
The energy that an electron gains by tunneling through the $k$-th junction, 
$\Delta E_k^{\leftarrow}$ or $\Delta E_k^{\rightarrow}$, 
may be calculated by integrating the 
difference between the neighboring Fermi levels over the tunneling 
event:  
\begin{equation}
\left. {\Delta E_k^{\leftarrow} \atop
\Delta E_k^{\rightarrow}} \right\}=
-e \int\limits_0^{\pm 1} dq ~
V_k(\ldots,n_{k-1}-q,n_k+q,\ldots,V).
\end{equation}
This is just the change in the electrostatic energy of the $k$-th capacitor.
Evaluating the integral leads to a simple expression for the energy
changes:
\begin{eqnarray}\label{deltaE}
\left. {\Delta E_k^{\leftarrow} \atop
\Delta E_k^{\rightarrow}} \right\}=
\mp e V_k(\ldots n_j\ldots,V) 
 \frac{e^2}{2} ~ \frac{C_1}{C_k^2 (1+C_1 \sum\limits_{m=2}^N 1/C_m)} 
- \frac{e^2}{2 C_k}. 
\end{eqnarray}

The $I(V)$ characteristics of this system can be determined numerically
by Monte Carlo simulation.
One main feature of the results which arises in the 
simulations is a widening of the Coulomb gap with increasing number
of tunnel junctions (Fig.\nolinebreak \ref{iv_3_5_10}).
This can be 
understood for the zero-temperature limit of the system. Also, the
differential conductivity  at  current onset (see inset of Fig.\nolinebreak
\ref{step_slope3}) can be calculated in
this limit.
For $T=0$, a tunneling event is impossible
whenever a tunneling electron would lose energy, that is 
$\Delta E_k^{\leftarrow}<0$ or $\Delta E_k^{\rightarrow}<\nolinebreak 0$, respectively.
From any state defined by the numbers $n_j$ of extra electrons 
on the center electrodes, 
the system may undergo a number of transitions by tunneling of an electron
somewhere in the series. 
The energy changes associated with these transitions,
$\Delta E_k^{\leftarrow}$ and $\Delta E_k^{\rightarrow}$, are 
directly related to the applied voltage via Eq.\nolinebreak (\ref{deltaE}).
In the simplest case of only one extra electron in the series
(or one extra hole, i.e., the absence of one electron),
these energy differences are negative for zero voltage, i.e., 
tunneling is impossible. 
An increasing voltage causes the energies to increase towards and then
above zero, thus increasing the number of possible tunneling events.
The energy differences associated with tunneling through
the terminating junctions of the chain are the last 
to become positive.
Therefore, the Coulomb gap of multiple junction systems
can be determined by examining the voltages at which either 
$\Delta E_{1}^{\rightarrow}$ and $\Delta E_{1}^{\leftarrow}$ or
$\Delta E_{N}^{\rightarrow}$ and  $\Delta E_{N}^{\leftarrow}$ cross zero. 
For $C_1<C_N$, 
the first tunnel junction is the one that dominates the process and
the Coulomb gap is found to be 

\begin{equation}
\label{gap}
\Delta V_{\scriptsize{\mbox{gap}}}=e\sum\limits_{i=2}^{N}\frac{1}{C_i}
\end{equation}
For $C_1>C_N$, the last junction dominates the process, which
means that the sum in Eq.\nolinebreak (\ref{gap}) runs from $i=1$ to $N-1$.
A similar approach to calculate the size of the Coulomb
gap was chosen before,\cite{laikhtman}
although the system of junctions is not as general as that considered 
here (the capacitances are taken to be all equal).

Higher threshold voltages can in principle
be determined in a similar manner,
 thus accounting for the structure of multiple
junction $I(V)$ curves. 
Using a fairly simple computer program, one can calculate
the states that are accessible from the 
``ground state'', which is given by $n_j=0$ for all $j$'s,
and one can 
determine the threshold voltages at which new states become accessible. 
A state is accessible if a sequence of tunneling events with nonvanishing 
probability leads to it.
These threshold voltages provide an intuitive understanding of
the steps in the $I(V)$ characteristic as well as an exact prediction of
the voltages at which they occur.
Until now, the step structures in such simulations have usually been
presented without explanation or simply been characterized as
unusual.\cite{barsadeh,amman}
In the present analysis, however, 
the number of predicted threshold voltages exceeds the number of steps that 
occur in the numerical simulation. The occurence of steps depends
crucially on the choice of the resistance values, as will be discussed
subsequently.

Just beyond the Coulomb gap, i.e., 
at voltages slightly above the first threshold voltage 
$V_{\scriptsize{\mbox{\scriptsize th}}}=
\frac{1}{2}\Delta V_{\scriptsize{\mbox{gap}}}$, 
the most probable process is one electron (hole) tunneling through
the whole chain of junctions before the next electron (hole) enters. 
The average time that the electron spends on one particle of the 
chain is given by the inverse tunneling rate.
The average time $\tau$ to tunnel through the whole chain is therefore
\begin{eqnarray}
\label{time}
\tau = \frac{1}{l_{N}(0,\ldots,0,V)}+\frac{1}{l_{N-1}(0,\ldots, 0,1,V)} 
+\ldots
+\frac{1}{l_{1}(1,0,\ldots,0,V)}.
\end{eqnarray}
This time $\tau$ is valid for voltages $V$ slightly above the threshold
voltage: $V=V_{\mbox{\scriptsize th}}+\delta V$, where $\delta V$ is small. 
At zero temperature, the
tunneling rates $l_k$ are given by the expresssion
\begin{equation}
l_k(\ldots n_j\ldots, V)=
\left\{
\begin{array}{cl}
0 & \mbox{ for }\Delta E_k^{\rightarrow}< 0,\\
 \Delta E_k^{\rightarrow}/(e^2 R_k) & \mbox{ for }\Delta E_k^{\rightarrow}\geq 0.
\end{array}
\right.
\end{equation}
Therefore, just above the threshold voltage $V_{\mbox{\scriptsize th}}$, 
the voltage at which the energy change at the $N$th junction,
\[\Delta E_{N}^{\rightarrow}=\frac{e}{C_N\ \sum\limits_{m=1}^{N} 1/C_m}
~\delta V,\] 
crosses zero,
the first term in Eq.\nolinebreak (\ref{time}) 
dominates. 
The $I(V)$
dependence near $V_{\mbox{\scriptsize th}}$ is thus given by
\begin{eqnarray}
\label{ivnearvth}
I(V)&=&\frac{e}{\tau} 
=e\ l_N(0,\ldots,0, V) \nonumber \\
&=&
\left\{
\begin{array}{cl}
0&\ \mbox{for } V<V_{\mbox{\scriptsize th}},\\
\left(C_N R_N\right. 
\sum\limits_{m=1}^{N} \left. 1/C_m\right)^{-1}
\delta V&\ \mbox{for } V \geq V_{\mbox{\scriptsize th}}.
\end{array}
\right.
\end{eqnarray}

In the inset of Fig.\nolinebreak \ref{step_slope3}, Monte Carlo results
obtained close to the first threshold voltage
are shown for two sets of parameters 
and are compared with the asymptotes given by Eq.\nolinebreak (\ref{ivnearvth}).
This figure demonstrates how drastically the differential conductivities
at current onset, given the same average slope
of $1/\sum_{i=1}^N R_i$, depend on the value of the resistance of 
junction 1, which in this case determines the threshold voltage.
It also shows the good agreement 
of the numerical data with the analytical result.

In a more rigorous approach,
probabilities $P(\ldots n_j \ldots)$ must be assgined to the
states $(\ldots n_j \ldots)$ of the system. These probabilities
must be solutions of
the stationary master equation of the system.
The correct expansion for the stationary current is then given by
\begin{eqnarray}
\label{correctexpansion}
I &=& e\sum\limits_{\scriptsize\mbox{all~}
(\ldots n_j\ldots)}P(\ldots n_j\ldots) 
\left[l_N(\ldots n_j\ldots, V)-r_N(\ldots n_j\ldots,
V)\right],
\end{eqnarray}
summing over all possible electronic states $(\ldots n_j\ldots)$ of the system
with an arbitrary number of extra electrons on the center electrodes.

For voltages $V=V_{\mbox{\scriptsize th}}+\delta V$ near the first 
threshold voltage, the probabilty of the ``ground
state'' $(0,\ldots ,0)$
is almost 1, i.e., $P(0,\ldots ,0)\simeq 1$, and the probabilities of all 
other states are negligible. This results from the fact that, while 
the transition rates 
from the state $(0,\ldots ,0)$ to neighboring states are zero
or nearly zero, 
the transition rates back to the ground
 state are already finite. Furthermore,
$r_N(0,\ldots ,0, V)$ vanishes for positive voltages $V$, leaving only 
$l_N(0,\ldots,0,V)$.
Therefore Eq.\nolinebreak (\ref{correctexpansion}) reduces to
\begin{equation}
I = e\ l_N(0,\ldots,0, V)
\end{equation} 
near the first threshold voltage, which is the same result as
obtained above in Eq.\nolinebreak (\ref{ivnearvth}).

It is, of course, quite difficult to calculate analytically
the probabilties of the states from the full master equation of the
system, although it is possible (for capacitances all equal)
to write an analytical expression for the 
average current without an explicit solution
of the master
equation in the case when the voltage corresponds to the first two steps
in the $I(V)$ characteristic.\cite{korotkov} 
Qualitatively, one can  expect --- as at the first
threshold voltage ---
the steepest steps in $I(V)$ to occur when
tunneling through the junction with the lowest resistance value
creates access to a new state.
Higher steps may only occur 
when the associated threshold voltage is
determined by a junction $k_0$ for which
\[
\left(C_{k_0}~R_{k_0}\right. \sum\limits_{m=1}^{N} \left. 1/C_m\right)^{-1}\gg
\frac{1}{\sum\limits_{i=1}^{N}R_i}.
\]

This tendency can also be seen in the results of Fig.\nolinebreak \ref{step_slope3}. 
The solid arrows indicate voltages 
at which new states become accessible by tunneling through
junction 1, which manifests itself as steps in the (thicker) 
$I(V)$ curve for parameter 
set 1 ($R_1$ small). At voltages indicated by the dashed arrows, 
new states become accessible by tunneling through the last junction 6,
corresponding to steps in the (thinner)
curve for parameter set 2 ($R_6$ small). 

In conclusion, Monte Carlo simulations of $I(V)$ characteristics in a
one-dimensional arrangement
of tunnel junctions as well as an analytical approach have been presented.
The analytical results 
may be used to understand the features found in
the simulations.
The results are in general useful for the qualitative and quantitative
understanding of experimental $I(V)$ curves.
The voltages at which steps occur can be predicted
using the presented formalism --- in particular, the
size of the Coulomb gap can be determined.
Efforts are currently in progress to extend the theoretical approach to finite
temperatures and to investigate in more detail the influence of certain
system parameters, such as the resistance values in the junction array, on the
resulting $I(V)$ curves.
\acknowledgements
This work was supported by BMBF (Grant: 13N6562).


\begin{figure}
\caption{\label{schalt3}
(a) Circuit diagram of a system of $N$ tunnel junctions in series,
represented by their capacitances $C_i$. 
The voltage drop across the $i$-th tunnel junction is $V_i$.
(b) Schematic diagram of the tunnel junctions coupled in series, 
showing the various tunneling rates $l_i$, $r_i$.
}
\end{figure}

\begin{figure}
\caption{\label{iv_3_5_10}
Monte Carlo simulations of $I(V)$ curves for (a) 3, (b) 5 and (c) 10 tunnel 
junctions. 
The parameters for the simulations are $C_1=1.2$ aF, 
$C_2=\cdots=C_{N-1}=1.4$ aF,  $C_N=2.8$ aF and
$R_1=0.01\mbox{ M}\Omega$,
$R_2=\cdots=R_{N-1}=280{ M}\Omega$,  $R_N=2.8\mbox{ M}\Omega$ and
$T=4.2$K.
}
\end{figure}

\begin{figure}
\caption{\label{step_slope3}
Monte Carlo simulation and analytical predictions for the $I(V)$ 
step structure
of a 6-junction system. The parameters are $T=0.01$K, $C_1=1.2$ aF, 
$C_{2\ldots 5}=2.8$ aF, $C_1=4.4$ aF (in both cases),
 $R_1=24$M$\Omega$, $R_{2\ldots 5}=560$M$\Omega$, $R_6=8000$M$\Omega$ 
(thick line, set 1) and 
$R_1=8000$M$\Omega$, $R_{2\ldots 5}=560$M$\Omega$, $R_6=24$M$\Omega$ (thin
line, set 2). Solid arrows indicate access through junction 1,
dashed arrows through junction 6 (see text). 
In the inset,
current onsets for the two sets of parameters are magnified.
The diamonds (set 1) and squares (set 2)
are Monte Carlo results, whereas the solid lines
display the asymptotes for $V\rightarrow V_{\mbox{th}}$ from the analytical
expression. The dot-dashed line shows the average slope for
reference.
}
\end{figure}

\end{document}